\documentstyle[11pt,epsf,epsfig]{article}

\def\midformat{
\setlength{\textheight}{8.9in} \setlength{\textwidth}{6.7in}
\setlength{\evensidemargin}{-0.19in}
\setlength{\oddsidemargin}{-0.19in} \setlength{\headheight}{0in}
\setlength{\headsep}{10pt} \setlength{\topsep}{0in}
\setlength{\topmargin}{0.0in}
\setlength{\itemsep}{0in}       
\renewcommand{\baselinestretch}{1.1}
\parskip=0.070in
}

\newtheorem{theorem}{Theorem}[section]

\newtheorem{claim}[theorem]{Claim}
\newtheorem{lemma}[theorem]{Lemma}

\newtheorem{comment}[theorem]{Comment}

\def\boldhead#1:{\par\vskip 7pt\noindent{\bf #1:}\hskip 10pt}
\def\ithead#1:{\par\vskip 7pt\noindent{\it #1:}\hskip 10pt}

\def\inline#1:{\par\vskip 7pt\noindent{\bf #1:}\hskip 10pt}
\long\def\comment #1\commentend{} \long\def\commhide
#1\commhideend{} \long\def\commfull #1\commend{#1}
\long\def\commabs #1\commenda{} \long\def\commtim #1\commendt{#1}
\long\def\commb #1\commbend{}
\long\def\comln #1\comlnend{#1}   
\long\def\comex #1\comexend{}     

\long\def\CProof #1\CQED{}
\def\blackslug{\hbox{\hskip 1pt \vrule width 4pt height 8pt
    depth 1.5pt \hskip 1pt}}
\def\QED{\quad\blackslug\lower 8.5pt\null\par}

\long\def\PPP#1{\noindent{\bf Proof:}{ #1}{\quad\blackslug\lower
8.5pt\null}}

\long\def\denspar #1\densend {#1}

\def\cD{{\cal D}}

\def\cF{{\cal F}}
\def\cG{{\cal G}}

\def\cM{{\cal M}}

\def\cP{{\cal P}}

\def\cR{{\cal R}}

\def\cT{{\cal T}}

\newif\ifnotesw\noteswtrue
   {\ifnotesw\marginpar[\hfill\(\top\)]{\(\top\)}\fi}%
   {\ifnotesw\marginpar[\hfill\(\bot\)]{\(\bot\)}\fi}

\newcommand{\mnote}[1]%
    {\ifnotesw\marginpar%
        [{\scriptsize\it\begin{minipage}[t]{\marginparwidth}
        \raggedleft#1%
                        \end{minipage}}]%
        {\scriptsize\it\begin{minipage}[t]{\marginparwidth}
        \raggedright#1%
                        \end{minipage}}%
    \fi}
\def\MathF{\hbox{\rm I\kern-2pt F}}
\def\MathP{\hbox{\rm I\kern-2pt P}}
\def\MathR{\hbox{\rm I\kern-2pt R}}
\def\MathZ{\hbox{\sf Z\kern-4pt Z}}
\def\MathN{\hbox{\rm I\kern-2pt I\kern-3.1pt N}}
\def\MathC{\hbox{\rm \kern0.7pt\raise0.8pt\hbox{\footnotesize I}
\kern-4.2pt C}}

\newcommand{\ignore}[1]{}

\midformat

\newcommand{\MC}{{\cal MC}}

\begin{document}

\title{Labeling Schemes with Queries\\}
\author{
Amos Korman
\thanks{Information Systems Group, Faculty of IE\&M, The Technion, Haifa, 32000
Israel.  E-mail:{\tt pandit@tx.technion.ac.il}. Supported in part at
the Technion by an Aly Kaufman fellowship.} \and Shay Kutten
\thanks{Information Systems Group, Faculty of IE\&M, The Technion,
Haifa, 32000 Israel.  E-mail: {\tt kutten@ie.technion.ac.il}.
 Fax: +972 (4) 829 5688. Phone: +972 (4) 829 4505. Supported in part
by a grant from the Israeli Ministry for Science and Technology.}}

\begin{titlepage}
\def\thepage{}
\maketitle

\begin{abstract}
We study the question of ``how robust are the known lower bounds of
labeling schemes when one increases the number of consulted
labels''. Let $f$ be a function on pairs of vertices. An
$f$-labeling scheme for a family of graphs $\cF$ labels the vertices
of all graphs in $\cF$ such  that for every graph $G\in\cF$ and
every
 two vertices $u,v\in
G$, the value $f(u,v)$ can be inferred by merely inspecting the
labels of $u$ and $v$.

This paper introduces a natural generalization: the notion of
$f$-labeling schemes with queries, in which the value  $f(u,v)$ can
be inferred by inspecting not only the labels of $u$ and $v$ but
possibly the labels of some additional vertices. We show that
inspecting the label of a single additional vertex (one {\em query})
enables us to reduce the label size of many labeling schemes
significantly. In particular, we show that to support the distance
function on $n$-node trees as well as the flow function on $n$-node
general graphs, $O(\log n+\log W)$-bit labels are sufficient and
necessary, where $W$ is the maximum (integral) capacity of an edge.
We note that it was shown that any labeling scheme (without queries)
supporting either the flow function on general graphs or the
distance function on trees, must have label size $\Omega(\log^2
n+\log n\log W)$. Using a single query, we also show a routing
labeling scheme in the fixed-port model using $O(\log n)$-bit
labels, while the lower bound on the label size of any such routing
labeling scheme (without queries) is $\Omega({\log^2 n}/{\log\log
n})$.  We note that all our labeling schemes with queries have
asymptotically optimal label size.

We then show several extensions. In particular, we show that our
labeling schemes with queries on trees can be extended to the
dynamic scenario by some adaptations of known model translation
methods. Second, we show that the study of the queries model can
help with the traditional model too.  That is, using ideas from
our routing labeling scheme with one query, we show how to
construct a 3-approximation routing scheme {\em without queries}
in the fixed-port model with $\Theta(\log n)$-bit labels. Finally,
we turn to a non-distributed environment and first translate known
results on finding NCA to supporting distance queries. We then use
the above to show that one can preprocess a general weighted graph
using almost linear space so that flow queries can be answered in
almost constant time.

\end{abstract}

\end{titlepage}
\pagenumbering{arabic}

\section{Introduction}
\label{sec:introduction}
\paragraph*{\bf Background:}
Network representations play a major role in many domains of
computer science, ranging from data structures, graph algorithms,
and combinatorial optimization to databases, distributed computing,
and communication networks. In most traditional network
representations, the names or identifiers given to the vertices
betray no useful information, and they serve only as pointers to
entries in the data structure, which forms a {\em global}
representation of the network. Recently, quite a few papers studied
methods for representing network properties by assigning {\em
informative labels} to the vertices of the network (see e.g.,
\cite{KNR92,AR02+,GPPR01,KKKP04,Peleg00:lca,AGKR01,FG01}).

Let $f$ be a function on pairs of vertices (e.g., distance).
Informally, the goal of an $f$-labeling scheme is to label the
vertices of a graph $G$ in such a way  that for every two vertices
$u,v\in G$, the value $f(u,v)$ (e.g., the distance between $u$ and
$v$) can be inferred by merely inspecting
the labels of $u$ and $v$.

Of course, this can be done trivially using labels that are large
enough (e.g., every label includes the description of the whole
graph). Therefore, the main focus of the research concerning
labeling schemes is to minimize the amount of information (the sizes
of the labels) required. Informally, an $f$-labeling scheme can be
viewed as a way of distributing the graph structure information
concerning $f$ to the vertices of the graph, using small chunks of
information per vertex.

This paper introduces the notion of $f${\em-labeling schemes with
queries} which  generalizes  the notion of $f$-labeling schemes. The
idea is to distribute the global information (relevant to $f$) to
the vertices, in such a way that $f(u,v)$ can be inferred by
inspecting not only the labels of $u$ and $v$ but possibly the
labels of some additional vertices. We note that all the
constructions given in this paper calculate $f(u,v)$ by inspecting
 the labels of three vertices ($u$ and $v$ above, and some $w$).
That is, given the labels of $u$ and $v$, we first find a vertex $w$
and then consult its label to derive $f(u,v)$.

Our paper demonstrate that the task of distributing the graph
structure to the vertices using short labels is very sensitive to
the model used. In particular, in order to calculate $f(u,v)$, if
 one is allowed to consult the labels of three vertices instead of two,
 then the sizes of the labels can drop considerably, in orders of magnitude.
Moreover, our query labeling schemes can be obtained using simple
methods. This turns out to be useful, when we go back to the
traditional model (without queries) and apply ideas from one of our
methods to construct a new result in the traditional model.

Most previous research concerning distributed network
representations considered the {\em static} scenario, in which the
topology of the underlying network is fixed. However, recent
papers tackle the more difficult task of labeling dynamic networks
in a distributed fashion. Such labeling methods should, of course,
be dynamic too. We show that the effect of introducing a query to
the dynamic case is similar to the effect on the static case. That
is, we modify the model translation methods of \cite{K05} and
\cite{KPR04}, and then use them to extend our static labeling
schemes with queries on trees, to the dynamic scenario. We then
show that the sizes of the resulted schemes are smaller than those
of the schemes for the older model. The reduction in the label
size is similar to the reduction in the static case.

Finally, we show that our methods are also useful in
the non-distributed environment.

\paragraph*{Related work:}
In this subsection we mostly survey results concerning labeling
schemes (with no queries). However, let us first mention an area of
research (namely, overlay and Peer to Peer networks) that may serve
as a practical motivation for our work, and for some other studies
concerning labeling schemes. We stress, though, that the main
motivation for this paper is theoretical.

When the third vertex $w$ (mentioned above) is near by to $u$, it
may be quite cheap for $u$ to access the main memory at $w$,
sometimes even cheaper than consulting the disk at $u$ itself. See,
for example \cite{ercim,subpages}. Indeed, some of our schemes below
are based on such a ``near by'' $w$. Even when $w$ is remote,
accessing it may be cheap in some overlay networks. The main
overhead there is finding $w$ (which can be done in our
constructions by $u$ using $v$'s label) and creating the connection
to it. Such models are presented explicitly, e.g. in
\cite{HLL,CGK,AACWY,CAN}, where such remote accesses are used to
construct and to use overlay data structures. Famous overlay data
structures that can fit such models appear for example in
\cite{Chord,tapestry,viceroy,debrujin-pierre,debrujin-udi}. In some
of these overlay networks, a vertex $w$ is addressed by its
contents. This may motivate common labeling schemes that assume
content addressability.

Implicit labeling schemes  were first introduced in
\cite{Breuer66,BF-67,KNR92}. Labeling schemes supporting the
adjacency and ancestry functions on trees were investigated in
\cite{KNR92,AR02+,KM01,AKM01,AR02}.

Distance labeling schemes were studied in
\cite{P99:lbl,KKP00,GPPR01,GKKPP01,GP01a,KM01,T01,CHKZ02,ABR03}. In
particular, \cite{P99:lbl} showed that the family of $n$ vertex
weighted trees with integer edge capacity of at most $W$ enjoys a
scheme using $O(\log^2 n+\log n\log W)$-bit labels. This bound was
proven in \cite{GPPR01} to be asymptotically optimal.

Labeling schemes for routing on trees were investigated in a number
of papers \cite{SK85,VT87,C99,EGP96,G00} until finally optimized in
\cite{FG01,FG02,TZ01}. For the {\em designer port} model, in which
the designer of the scheme can freely enumerate the port numbers of
the nodes, \cite{FG01} shows how to construct a routing scheme using
labels of $O(\log n)$ bits on $n$-node trees. In the {\em adversary
port} model, in which the port numbers are fixed by an adversary,
they show how to construct a routing scheme using labels of
$O(\log^2 n/\log\log n)$ bits on $n$-node trees. In \cite{FG02} they
show that both label sizes are asymptotically optimal.
Independently, a  routing  scheme for trees using $(1+o(1))\log
n$-bit labels was introduced in \cite{TZ01} for the designer port
model.

Two variants of labeling schemes supporting the nearest common
ancestor (NCA) function in trees appear in the literature. In an
id-NCA labeling schemes, the vertices of the input graph are assumed
to have disjoint identifiers (using $O(\log n)$ bits) given by an
adversary. The goal of an id-NCA labeling scheme is to label the
vertices  such  that given the labels of any two vertices $u$ and
$v$, one can find the identifier of the NCA of $u$ and $v$. Static
labeling schemes on trees supporting the separation level and id-NCA
functions were given in \cite{Peleg00:lca} using $\Theta(\log^2
n)$-bit labels. The second variant considered is the label-NCA
labeling scheme, whose goal is to label the vertices such that given
the labels of any two vertices $u$ and $v$, one can find the label
(and not the pre-given identifier) of the NCA of $u$ and $v$. In
\cite{AGKR01} they present a label-NCA labeling scheme on trees
enjoying $\Theta(\log n)$-bit labels.

In \cite{KKKP04} they give a labeling scheme supporting the flow
function on $n$-node general graphs using $\Theta(\log^2 n+\log
n\log W)$-bit labels, where $W$ is the maximum capacity  of an edge.
They also show a labeling scheme supporting the
$k$-vertex-connectivity function on general graphs using $O(2^k\log
n)$-bit labels. See \cite{GP01b} for a survey on (static) labeling
schemes.

Most of the research concerning labeling schemes in the dynamic
settings considered the following two dynamic models on tree
topologies. In the {\em leaf-dynamic} tree model, the topological
event that may occur is that a leaf is either added to or removed
from the tree. In
 the {\em leaf-increasing} tree model, the only topological event that may
occur is that a leaf joins the tree.

A labeling schemes for the ancestry relation in the leaf-dynamic
tree model is given in \cite{CKM02} using $O(n)$-bit labels. They
assume that once a label is given to a node it remains unchanged.
Therefore, the issue of updates is not considered even for the non
distributed setting. Other labeling schemes are presented in the
above paper assuming that clues about the future topology of the
dynamic tree are given throughout the scenario.

The study of dynamic distributed labeling schemes was initiated in
\cite{KPR04,KP03}. In \cite{KPR04}, a dynamic  labeling scheme is
presented for distances in the leaf-dynamic tree model with
$O(\log^2 n)$ label size and $O(\log^2 n)$ amortized message
complexity, where $n$ is the current tree size.
$\beta$-approximate distance labeling schemes (in which, given two
labels, one can infer a $\beta$-approximation to the distance
between the corresponding nodes) are presented \cite{KP03}.
 Their schemes apply for dynamic models in which the tree
topology is fixed but the edge weights may change.

Two general translation methods for extending static labeling
schemes on trees to the dynamic setting are considered in the
literature. Both approaches fit a number of natural functions on
trees, such as ancestry, routing, label-NCA, id-NCA  etc.  Given a
static labeling scheme on trees, in the leaf-increasing tree
model, the resulting dynamic scheme in \cite{KPR04} incurs
overheads (over the static scheme) of $O(\log n)$ in both the
label size and the communication complexity. Moreover, if an upper
bound $n_f$ on the final number of vertices in the tree is known
in advance, the resulting dynamic scheme in \cite{KPR04} incurs
overheads (over the static scheme) of $O(\log^2 n_f / \log\log
n_f)$ in the label size and only $O(\log n / \log\log n)$ in the
communication complexity. In the leaf-dynamic tree model there is
an extra additive factor of $O(\log^2 n)$ to the amortized message
complexity of the resulted schemes.

In \cite{K05}, it is shown how to construct for many functions
$k(x)$, a dynamic labeling scheme in the leaf-increasing tree model
extending a given static scheme, such that the resulting scheme
incurs overheads (over the static scheme) of $O(\log_{k(n)} n)$ in
the label size and $O(k(n)\log_{k(n)} n)$ in the communication
complexity. As in \cite{KPR04}, in the leaf-dynamic tree model there
is an extra additive factor of $O(\log^2 n)$ to the amortized
message complexity of the resulted schemes. In particular, by
setting $k(n)={n^{\epsilon}}$, dynamic labeling schemes are obtained
with the same asymptotic label size as the corresponding static
schemes and sublinear amortized message, namely, $O( n^{\epsilon})$.

\subsection{Our contribution}
We introduce the notion of $f$-labeling schemes with queries that is
a natural generalization of the notion of $f$-labeling schemes.
Using this notion we demonstrate that by increasing slightly the
number of vertices whose labels are inspected, the size of the
labels decreases considerably.
Specifically, we inspect the labels of 3 vertices instead of 2, that
is, we use a single {\em query}.
In particular, we show that there exist
simple labeling schemes with one query supporting the distance
function on $n$-node trees as well as the flow function on $n$-node
general graphs with label size $O(\log n+\log W)$, where $W$ is the
maximum (integral) capacity of an edge. (We note
that the lower bound for labeling schemes without queries for each of
these problems is
$\Omega(\log^2 n+\log n\log W)$
\cite{GPPR01,KKKP04}.)
We also show that there exists a labeling scheme
with one query supporting the
 id-NCA function on $n$-node trees with label size
$O(\log n)$. (The lower bound for schemes without queries is
$\Omega(\log^2 n)$ \cite{Peleg00:lca}.) In addition, we show a
routing labeling scheme with one query in the fixed-port model using
$O(\log n)$-bit labels, while the lower bound (see \cite{FG02}) for
the case of no queries is $\Omega(\frac{\log^2 n}{\log\log n})$. We
note that all the schemes we introduce have asymptotically optimal
label size for schemes with one query. The matching lower bound
proofs are straightforward in most of the cases, and we present the
proofs in the remaining cases. Moreover, most of the results are
obtained by simple constructions, which strengthens the motivation
for this model.

We then show several extensions that are somewhat more involved. In
particular, we show that our labeling schemes with queries on trees
can be extended to the dynamic scenario using model translation
methods based on those of \cite{KPR04,K05}. In order to save in the
message complexity, we needed to make some adaptations to those
methods, as well as to one of the static routing schemes of
\cite{FG01}. Second, we show that the study of the queries model can
help with the traditional model too. That is, using ideas from our
routing labeling scheme with one query, we show how to construct a
3-approximation routing scheme {\em without queries} for unweighted
trees in the fixed-port model with $\Theta(\log n)$-bit labels.

Finally, we turn to a non-distributed environment and demonstrate
similar constructions. That is, first, we show a simple method to
transform previous results on NCA queries on static and dynamic
trees in order to support also distance queries. Then, we show that
one can preprocess a general weighted graph using almost linear
space so that flow queries can be answered in almost constant time.

\section{Preliminaries}
\label{sec:preliminaries}
 Let $T$ be a tree and let $v$ be a
vertex in $T$. Let $deg(v)$ denote the degree of $v$. For a non-root
vertex $v\in T$, let $p(v)$ denote the parent of $v$ in $T$. In the
case where the tree $T$ is weighted (respectively, unweighted), the
{\em depth} of a vertex is defined as its weighted (resp.,
unweighted) distance to the root. The {\em nearest common ancestor}
of $u$ and $w$, $NCA(u,w)$, is the common ancestor of both $u$ and
$w$ of maximum depth. Let $\cT(n)$ denote the family of all $n$-node
unweighted trees. Let $\cT(n,W)$ (respectively, $\cG(n,W)$) denote
the family of all  $n$-node weighted trees (resp., connected graphs)
with (integral) edge weights bounded from above by $W$.

Incoming and outgoing links from every node are identified by so
called {\em port-numbers}. We distinguish between the following two
variants of port models regarding routing schemes. In the {\em
designer port} model the designer of the scheme can freely assign
the port numbers of each vertex (as long as these port numbers are
unique), and in the {\em fixed-port} model the port numbers at each
vertex are assigned by an adversary. We assume that each port number
is encoded using $O(\log n)$ bits.
\subsection{The functions}
We consider the following functions which are applied on pairs of
vertices $u$ and $v$ in a graph $G=\langle V,E\rangle$. The
detailed definitions
of these functions are given in the appendix.\\
(1) {\bf flow} (maximum legal flow between $u$ and $v$), (2) {\bf
distance} (graph distance, either weighted, or unweighted), (3) {\bf
routing} (the port in $u$ to the next vertex towards $v$). If the
graph is a tree $T$ then we consider also the following functions:
(4) {\bf separation level} (depth of $NCA(u,v)$), (5) {\bf id-NCA},
(6) {\bf label-NCA}. In (5) above, it is assumed that
 identities containing $O(\log
n)$ bits are assigned to the vertices by an adversary, and
 $id-NCA_T(u,v)$ is the identity of $NCA(u,v)$. In (6) above, it is
 assumed
that each vertex can freely select its own identity (as long as all
identities remain unique). In this case, the identities may also be
referred to as labels.)

\subsection{Labeling schemes and $c$-query labeling schemes}
\label{def}
Let $f$ be a function defined on pairs of vertices of a graph. An
{\em $f$-labeling scheme} $\pi=\langle \cM,\cD \rangle$ for a family
of graphs $\cF$ is composed of the following components:
\begin{enumerate}
\item A {\em marker} algorithm $\cM$ that given a graph $G\in\cF$,
assigns a label $\cM(v)$ to each vertex $v\in G$.
\item A (polynomial time) {\em
decoder} algorithm $\cD$ that given the labels $\cM(u)$ and $\cM(v)$
of two vertices $u$ and $v$ in some graph $G\in \cF$, outputs
$f(u,v)$.
\end{enumerate}

The most common measure used to evaluate a labeling scheme
$\pi=\langle \cM,\cD \rangle$, is the {\em label size}, i.e., the
maximum number of bits used in a label $\cM(v)$ over all vertices
$v$ in all graphs $G\in \cF$.

Let $c$ be some constant integer. Informally, in contrast to an
$f$-labeling scheme, in a  $c$-query $f$-labeling scheme, given
the labels of two vertices $u$ and $v$, the decoder may also
consult the labels of $c$ other vertices. More formally, a {\em
$c$-query $f$-labeling scheme} $\varphi=\langle \cM,Q,\cD \rangle$
is composed of the following components:
\begin{enumerate}
\item
A {\em marker} algorithm $\cM$ that given a graph $G\in\cF$, assigns
a label $\cM(v)$ to each vertex $v\in G$. This label is composed of
two sublabels, namely, $\cM^{index}(v)$ and $\cM^{data}(v)$, where
it is required that the index sublabels are unique, i.e., for every
two vertices $v$ and $u$, $\cM^{index}(v)\neq\cM^{index}(u)$.
(In other words, the index sublabels can serve as identities.)
\item
A (polynomial time) {\em query} algorithm $Q$ that given the labels
$\cM(u)$ and $\cM(v)$ of two vertices $u$ and $v$ in some graph
$G\in \cF$, outputs $Q(\cM(u),\cM(v))$ which is a set containing the
indices (i.e., the first sublabels) of $c$ vertices in $G$.
\item
A (polynomial time) {\em decoder} algorithm $\cD$ that given the
labels $\cM(u)$ and $\cM(v)$ of two vertices $u$ and $v$ and the
labels of the vertices in $Q(\cM(u),\cM(v))$, outputs $f(u,v)$.
\end{enumerate}

As in the case of $f$-labeling schemes, we evaluate a $c$-query
$f$-labeling scheme $\varphi=\langle \cM,Q,\cD \rangle$ by its
{\em label size}, i.e, the maximum number of bits used in a label
$\cM(v)$ over all vertices $v$ in all graphs $G\in \cF$. We note
that all the schemes in this paper use $c=1$, increasing the
number of labels used by a smallest constant, while the size of
the maximum label drops significantly in the order of magnitude.
Let us comment also that clearly, since the index sublabels must
be disjoint, any $c$-query $f$-labeling scheme on any family of
$n$-node graphs must have label size $\Omega(\log n)$. See Section
7 for alternative definition for query labeling schemes.

\subsection{Routing schemes and $\beta$-approximation routing schemes}
A {\em routing scheme} is composed of a {\em marker algorithm} $\cM$
for assigning each vertex $v$ of a graph $G$ with a label $\cM(v)$,
coupled with a {\em router} algorithm $\cR$ whose inputs are the
header of a message, $\cM(v)$ and the label $\cM(y)$ of a
destination vertex $y$. If a vertex $x$ wishes to send a message to
vertex $y$, it first prepares and attaches a header to the message.
Then the router algorithm $x$ outputs a port of $x$ on which the
message is delivered to the next vertex. This is repeated in every
vertex until the message reaches the destination vertex $y$. Each
intermediate vertex $u$ on the route may replace the header of the
message with a new header and may perform a local computation. The
requirement is that the weighted length of resulting path connecting
$x$ and $y$ is the same as the distance between $x$ and $y$ in $G$.

For a constant $\beta$, a $\beta$-approximation routing scheme is
the same as a routing scheme except that the requirement is that the
length of resulting route connecting $x$ and $y$ is a
$\beta$-approximation for the distance between $x$ and $y$ in $G$.

In addition to the label size, we also measure a routing scheme (and
a $\beta$-approximation routing scheme) by the {\em header size},
i.e., the maximum number of bits used in a header of a message.

\section{Labeling schemes with one query}
\label{sec:Labeling schemes with one query} In this section we
demonstrate that the query model allows for significantly shorter
labels. In particular, we describe simple 1-query labeling schemes
with labels that beat the lower bounds in the following well
studied cases: for the family of $n$-node trees, schemes
supporting the routing (in the fixed-port model), distance,
separation level, and the id-NCA functions; for the family of
$n$-node general graphs, a scheme supporting the flow function. We
note that all the schemes we present use asymptotically optimal
labels. Most of the 1-query labeling schemes obtained in this
section  use the label-NCA labeling scheme
$\pi_{NCA}=\langle\cM_{NCA},\cD_{NCA}\rangle$ described in
\cite{AGKR01}. Given an $n$-node, the marker algorithm $\cM_{NCA}$
assigns each vertex $v$ a distinct label $\cM_{NCA}(v)$ using
$O(\log n)$ bits. Given the labels $\cM_{NCA}(v)$ and
$\cM_{NCA}(u)$ of two vertices $v$ and $u$ in the tree, the
decoder $\cD_{NCA}$ outputs the label $\cM_{NCA}(w)$.

\subsection{Id-NCA function in trees}\label{id} We first describe a 1-query
labeling scheme
$\varphi_{id-NCA}=\langle\cM_{id-NCA},Q_{id-NCA},\cD_{id-NCA}\rangle$
that demonstrates how easy it is to support the id-NCA function on
$\cT(n)$ using one query and $O(\log n)$-bit labels. (Recall that
the lower bound on schemes without queries
 is $\Omega(\log^2 n)$
\cite{Peleg00:lca}.)

Informally, the idea behind $\varphi_{id-NCA}$ is to have the labels
of $u$ and $v$ (their first sublabels) be the labels given by the
label-NCA labeling scheme $\pi_{NCA}(v)$. Hence, they are enough for
the query algorithm to find the $\pi_{NCA}$ label of their nearest
common ancestor $w$. Then, the decoder algorithm finds $w$'s
identity simply in the second sublabel of $w$.

Let us now describe the 1-query labeling scheme $\varphi_{id-NCA}$
more formally. Given a tree $T$, recall that it is assumed that each
vertex $v$ is assigned a unique identity $id(v)$ by an adversary and
that each such identity is composed of $O(\log n)$ bits. The marker
algorithm $\cM_{id-NCA}$ labels each vertex $v$ with the label
$\cM_{id-NCA}(v)=\langle\cM_{id-NCA}^{index}(v),\cM_{id-NCA}^{data}(v)\rangle=\langle\cM_{NCA}(v),id(v)\rangle$.
Given the labels $\cM_{id-NCA}(v)$ and $\cM_{id-NCA}(u)$ of two
vertices $v$ and $u$ in the tree, the query algorithm $Q_{id-NCA}$
uses the decoder $\cD_{NCA}$ applied on the corresponding first
sublabels to output the sublabel
$\cM_{id-NCA}^{index}(w)=\cM_{NCA}(w)$, where $w$ is the NCA of $v$
and $u$. Given the labels $\cM_{id-NCA}(v)$, $\cM_{id-NCA}(u)$ and
$\cM_{id-NCA}(w)$ where $w$ is the NCA of $v$ and $u$, the decoder
$\cD_{id-NCA}$ simply outputs the second sublabel of $w$, i.e.,
$\cM_{id-NCA}^{data}(w)=id(w)$. The fact that $\varphi_{id-NCA}$ is
a correct 1-query labeling scheme for the id-NCA function on
$\cT(n)$ follows from the correctness of the label-NCA labeling
scheme $\pi_{NCA}$. Since the label size of $\pi_{NCA}(v)$ is
$O(\log n)$ and since the identity of each vertex $v$ is encoded
using $O(\log n)$ bits, we obtain that the label size of
$\varphi_{id-NCA}$ is $O(\log n)$. As mentioned before, since the
index sublabels must be disjoint, any query labeling scheme on
$\cT(n)$ must have label size $\Omega(\log n)$. The following lemma
follows.
\begin{lemma}
The label size of a 1-query id-NCA labeling scheme on $\cT(n)$ is
$\Theta(\log n)$.
\end{lemma}
\subsection{Distance and separation level in trees}\label{distance}
The above method can be applied for other functions. For example,
let us now describe 1-query labeling schemes $\varphi_{sep-level}$
and $\varphi_{dist}$ supporting the distance and separation level
functions respectively on $\cT(n,W)$. Both our scheme have label
size $\Theta(\log n+\log W)$. Recall that any labeling scheme
(without queries) supporting
 either the distance function or the separation level function on $\cT(n,W)$ must
 have size $\Omega(\log^2 n+\log n\log W)$,
\cite{GPPR01,Peleg00:lca}. We first show the following claim.
\begin{claim}
\label{lower-bound} Let $c$ be a constant. Any $c$-query labeling
scheme supporting either the separation level function or the
distance function on $\cT(n,W)$ must have label size $\Omega(\log
W+\log n)$.
\end{claim}
\begin{proof}
As mentioned before, any query labeling scheme  on $\cT(n)$ must
have label size $\Omega(\log n)$. We show the proof for the distance
function. Similar proof holds for the separation level function.
First note that we may assume that $W\geq ((c+2)\cdot n^{c})^2$,
otherwise, the lower bound trivially follows since $\Omega(\log
n)=\Omega(\log n+\log W)$. Let $\varphi=\langle\cM,Q,\cD\rangle$ be
any $c$-query labeling scheme supporting the distance function. Let
$P$ be an $n$-node path rooted at one of its end-nodes $r$. Let $v$
be the (only) child of $r$. For every $1\leq i\leq W$, let $P_i$ be
the path $P$ such that the edge $(r,v)$ has weight $i$ and all other
edges have weight 1. For every $1\leq i\leq W$ and every vertex
$u\in P_i$, let $L_i(u)$ denote the label given to $u$ by the marker
algorithm $\cM$ applied on $P_i$.
 For every $1\leq i\leq W$, let $S_i$ be the set of $c$ vertices
 given by the query algorithm  applied on the labels of $r$ and $v$
 in $P_i$, i.e.,
 $S_i=Q(L_i(r),L_i(v))$. Since there are at most $n^c$ sets of $c$
 vertices, there exists a set $X\subset\{1,2,\cdots W\}$ such that
 $|X|\geq W/n^c$ and for every $i,j\in X$, $S^i=S^j$.
Let $S$ denote the set of $c$ vertices such that for every $i\in X$,
$S^i=S$. For each $i\in X$, given the labels $L_i(r)$, $L_i(v)$ and
the labels of the vertices in $S$ assigned by $\cM$ applied on
$P_i$, the decoder outputs $i$, which is the distance between $r$
and $v$ in $P_i$. Since $|X|\geq W/((c+2)\cdot n^c)$, there must
exists a vertex in $u\in\{r,v\}\cup S$ such that the set
$\{L_i(u)\mid i\in X\}$ contains $\frac{W}{(c+2)\cdot n^c}$ values.
Therefore, there must exist an $i\in X$ such that $L_i(u)$ contains
$\log W-\log((c+2)\cdot n^c)>(\log W)/2$ bits. The claim
follows.\QED
\end{proof}

We now show how to construct 1-query labeling schemes supporting the
separation level and distance functions using similar method to the
one described in Subsection \ref{id}. Both schemes are based on
keeping the depth of a vertex in its data sublabel (instead of its
identity). The correctness of the 1-query labeling scheme supporting
the  distance function is based on the following equation.
\begin{equation}
d(v,u)=depth(v)+depth(u)-2\cdot depth(NCA(v,u)).
\end{equation}
The description of these schemes as well as the proof of the
following lemma is deferred to the appendix.
\begin{lemma}\label{sep+dist}
The label size of a 1-query labeling scheme supporting either the
separation-level or the distance function on $\cT(n,W)$ is
$\Theta(\log n+\log W)$.
\end{lemma}

\subsection{Routing in trees using one query}
\label{rout} As mentioned in before, any 1-query routing labeling
scheme on $\cT(n)$ must have label size $\Omega(\log n)$. In this
subsection, we establish a 1-query routing labeling scheme
$\varphi_{fix}$ in the fixed-port model using $O(\log n)$-bit
labels.

In \cite{FG01}, they give a routing  scheme $\pi_{des}=\langle
\cM_{des},\cD_{des}\rangle$
 for the designer port model in $\cT(n)$. Given a tree $T\in\cT(n)$, for every vertex
 $v\in T$, and every neighbor $u$ of $v$, let $port_{des}(v,u)$
denote the
 port number (assigned by the designer of the routing scheme $\pi_{des}$)
 leading from $v$ to $u$. In particular,
 the port number leading from each non-root vertex $v$ to its parent
$p(v)$ is assigned the number 1, i.e., $port_{des}(v,p(v))=1$. Given
the labels $\cM_{des}(v)$ and $\cM_{des}(w)$ of two vertices $v$ and
$w$ in $T$, the decoder $\cD_{des}$ outputs the port number
$port_{des}(v,u)$ at $v$ leading from $v$ to the next vertex $u$ on
the shortest path connecting $v$ and $w$.

Let $T$ be an $n$-node tree. We refer to a port number assigned by
the designer of the routing scheme $\pi_{des}$ as a {\em designer
port number} and to a  port number assigned by the adversary as an
{\em fixed-port number}. Let $port$ be some port of a vertex in
the fixed-port model. Besides having a fixed-port number assigned
by the adversary, we may also consider $port$ as having a designer
port number, the number that would have been assigned to it had we
been in the designer port model. For a port leading from vertex
$v$ to vertex $u$, let $port_{fix}(v,u)$ denote its fixed-port
number and let $port_{des}(v,u)$ denote its designer port number.

We now describe  our 1-query routing labeling scheme
$\varphi_{fix}=\langle \cM_{fix},Q_{fix},\cD_{fix}\rangle$ which
operates in the fixed-port model. Given a a tree $T\in\cT(n)$ and
a vertex $v\in T$, the index sublabel of $v$ is composed of two
fields, namely, $\cM^{index}(v)=\langle \cM^{index}_1(v),
\cM^{index}_2(v)\rangle$ and the data sublabel of $v$ is composed
of three fields, namely, $\cM^{data}(v)=\langle \cM^{data}_1(v),
\cM^{data}_2(v),\cM^{data}_3(v)\rangle$. If $v$ is not the root
then the index and data sublabels of $v$ are
$$\cM^{index}(v)=\langle \cM_{des}(p(v))~,~
port_{des}(p(v),v)\rangle~~~,~~~ \cM^{data}(v)=\langle
\cM_{des}(v)~,~ port_{fix}(p(v),v)~,~ port_{fix}(v,p(v)) \rangle.$$
Note that we use the designer port number as a part of the label in
the fixed-port model. Moreover, the designer port number at the
parent is used to label the child in the fixed-port model. Also note
that the index sublabel is unique, since $ \cM_{des}(x)$ must be
unique for $\pi_{des}$ to be a correct routing scheme.

The index sublabel of the root $r$ of $T$ is $\langle 0,0\rangle$
and the data sublabel of $r$ is $\cM^{data}(r)=\langle
\cM_{des}(r), 0,0\rangle$. Note that since the labels given by
the marker algorithm $\cM_{des}$ are unique, the index
sublabels of the vertices are unique.

Given the labels $\cM(v)$ and $\cM(w)$ of two vertices $v$ and
$w$, the decoder $\cD$ first checks whether
$\cD_{des}(\cM^{data}_1(v),\cM^{data}_1(w))=1$, i.e., whether the
next vertex on the shortest path leading from $v$ to $w$ is $v$'s
parent. In this case, the query algorithm is ignored and the
decoder $\cD_{fix}$ simply outputs $\cM^{data}_3(v)$ which is the
(fixed) port number at $v$ leading to its parent. Otherwise, the
query algorithm $Q_{fix}$ outputs $\langle
\cM^{data}_1(v),\cD_{des}(\cM^{data}_1(v),\cM^{data}_1(w))\rangle=\langle
\cM^{data}_1(v),\cD_{des}(\cM_{des}(v),\cM_{des}(w))\rangle$ which
is precisely  the index sublabel of $u$, the next vertex on the
shortest path leading from $v$ to $w$ (and a child of $v$), i.e.,
$\langle \cM^{data}_1(v),port_{des}(v,u)\rangle$. Therefore, given
labels $\cM_{fix}(v)$, $\cM_{fix}(w)$ and label $\cM_{fix}(u)$,
the decoder $\cD_{fix}$ outputs $\cM^{data}_2(u)$ which is the
desired port number $port_{fix}(v,u)$. Since the label size of
$\pi_{des}$ is $O(\log n)$ and since each port number is encoded
using  $O(\log n)$ bits, we obtain the following lemma.

\begin{lemma}
In the fixed-port model, the label size of a 1-query routing
labeling scheme on $\cT(n)$ is $\Theta(\log n)$.
\end{lemma}

\subsection{Flow in general graphs} We now consider the family
$\cG(n,W)$ of connected $n$-node weighted graphs with maximum edge
capacities $W$, and present a 1-query flow labeling scheme
$\varphi_{flow}$ for this family using $O(\log n+\log W)$-bit
labels. Recall that
 any labeling scheme
(without queries) supporting the flow function on $\cG(n,W)$ must
 have size $\Omega(\log^2 n+\log n\log W)$
\cite{KKKP04}.
The proof of the following lemma is sketched in the appendix.

\begin{lemma}
\label{lem:1qflow}
The label size of a 1-query flow labeling scheme on $\cG(n,W)$ is
$\Theta(\log n+\log W)$.
\end{lemma}

\section{3-approximation routing scheme in the fixed-port model
(without queries)}
By applying the method described in Subsection
\ref{rout} to the traditional model, we now show how to  construct a
3-approximation routing
 scheme (without queries) on $\cT(n)$. Our 3-approximation
routing labeling scheme $\pi_{approx}$ operates in the fixed-port
model and has label size and header size $O(\log n)$. Recall that
 any (precise) routing scheme
on $\cT(n)$ must have label size $\Omega(\log^2 n/\log\log n)$
\cite{FG02}. We note that our ideas for
 translating routing schemes from the designer port model to the
 fixed-port model implicitly appear in \cite{AGM04}, however, a 3-approximation routing
 scheme (without queries) on $\cT(n)$ is not explicitly constructed
 there. The description of the 3-approximation routing labeling scheme
$\pi_{approx}$ as well as the proof of the following lemma is
deferred to the Appendix.

\begin{lemma}\label{3-approx}
$\pi_{approx}$ is a correct 3-approximation routing scheme on
$\cT(n)$ operating in the fixed port model. Moreover, its label size
and header size are $\Theta(\log n)$.
\end{lemma}

\section{Adapting the 1-query schemes on trees to the dynamic setting}
\label{sec:adapting}

In this section we show how to translate our 1-query labeling
schemes on trees to the dynamic settings, i.e, to the
leaf-increasing and leaf-dynamic tree models, \cite{K05,KPR04} (see
also ``Related work'' in Section \ref{sec:introduction}). In a
dynamic scheme, the marker protocol updates the labels after every
topological change. We show that the reduction in the label sizes
obtained by introducing a single query in the dynamic scenario is
similar to the reduction in the static case.

To describe the adaptation fully, we need to give many details about
the methods of \cite{KPR04,K05,FG01}. Unfortunately, this is not
possible in this extended abstract. In our construction below, we
just state and then use some facts about these methods. These facts
can be proven easily given these methods.

 The initial idea is to apply the methods introduced in
\cite{KPR04,K05} to convert labeling schemes for static networks to
work on dynamic networks too. Unfortunately, we cannot do this
directly, since these methods were designed for traditional labeling
schemes and not for 1-query labeling schemes.

The next idea is to perform the conversion indirectly. That is,
recall (Section \ref{sec:Labeling schemes with one query}) that our
1-query labeling schemes utilize components that are schemes in the
traditional model (with no queries). That is, some utilize
$\pi_{NCA}$, the label-NCA labeling scheme of \cite{AGKR01} and some
utilize $\pi_{des}$, the routing scheme of \cite{FG01}. Hence, one
can first convert these components to the dynamic setting. Second,
one can attempt to use the resulted dynamic components in a similar
way that we used the static components in Section \ref{sec:Labeling
schemes with one query}. This turns out to be simple in the cases of
the distance, separation level and id-NCA functions, but more
involved in the case of the routing function.

 In the case of the distance and separation level functions,
 $\pi_{NCA}$ is used together with the depth of each vertex in its
label. Hence, in the dynamic case, we need to update the depth with
every topological change. In the dynamic models above, this uses
only a constant number of messages per topological change, since
once a vertex is added, its depth remains the same.

However, when trying to translate our 1-query routing labeling
scheme to the dynamic settings, things turn out to be more
difficult. Informally, the reason is the following. In
$\varphi_{fix}$, our static 1-query routing labeling scheme, each
non-root vertex `knows' the label assigned to its parent by another
static routing scheme $\pi_{des}$ (see Section \ref{sec:Labeling
schemes with one query}). When this scheme is made dynamic, the
natural translation that enables the child to continue to `know' the
label of its parent is the following:
\\
\noindent {\bf Notification}: whenever the label of a non-leaf
vertex changes, it notifies the new label to all the children. When
a child receives such a notification message, it updates its
$\varphi_{fix}$ label accordingly (see Section \ref{sec:Labeling
schemes with one query}).
\\

 In order to have an efficient translation, we need to account for
the messages used for the above notification. This turns out to be
relatively cheap when using the translation method of
\cite{KPR04}. The relevant facts about \cite{KPR04} is the
following: Theorem 4.16 and the fact that the label of a vertex
$v$ changes only as a part of changes in all the vertices of some
subtree $T'$. These changes in $T'$ involve $\Omega(|T'|)$
messages anyhow (not including the new notification messages
required from $v$ to its children). It so happened that in the
method of \cite{KPR04}, whenever a vertex $v$ is included in $T'$,
all its children are included in $T'$ as well. Hence, the cost of
the new notification messages can be amortized on the cost of the
messages sent anyhow in $T'$ to perform the changes. Hence, we can
(1) use the transformation of \cite{KPR04} to transform
$\pi_{des}$ to be dynamic, and then (2) use the new notification
messages to implement $\varphi_{fix}$ using the resulted dynamic
version of $\pi_{des}$. This yields a dynamic routing scheme with
1-query enjoying the complexities promised in \cite{KPR04} (see
part 2 in Theorem \ref{thm:dynamic} below). It is still left to
ensure parts 1 and 3 of  Theorem \ref{thm:dynamic} below. For that
we need to use the transformation of \cite{K05}, which turns out
to be less easy.

To prove parts 1 and 3 of Theorem \ref{thm:dynamic}, the relevant
fact about \cite{K05} consists of Theorems 1 and 2 of \cite{K05},
and of the list of the cases where messages are sent. The latter
is needed because we would like to show that we can amortize the
cost of the new messages we now send, on the messages sent anyhow
by the method of \cite{K05}. Unlike the case of \cite{KPR04}, here
messages are sent over a subtree that may include some vertex $v$,
but not all of its children. Recall that our modifications
includes the notification: sending messages from $v$, whose label
was changed, to all of its children. When $v$ has children outside
of $T'$, we cannot amortize the cost of the notification messages
on the cost of the messages sent anyhow by the method of
\cite{K05}. It turns out, however, that the only vertex in $T'$
such that not all of its children are necessarily in $T'$ is the
root $r'$ of $T'$. Therefore, the notifications messages sent by a
vertex $v\in T'$, where $v\neq r'$ do not increase the cost of the
messages sent anyhow by the method of \cite{K05}. The only thing
left to consider is the case where $r'$ changes its label in the
dynamic labeling scheme given by the method of \cite{K05} applied
on $\pi_{des}$, the static routing scheme of \cite{FG01}. To bound
the number of notification messages sent by  $r'$, we make sure
that the label of $r'$ changes only when $r'$ has at most 1 child.
This is done by modifying the static routing scheme $\pi_{des}$
such that when applying $\pi_{des}$ on any tree, the label given
by  $\pi_{des}$ to the root of the tree is always the same.

It therefore follows that the cost of the remaining notification
messages can be amortized on the cost of the messages sent anyhow by
\cite{K05}. The necessary modifications of the schemes, as well as
the proof of the following theorem, appear in the appendix.

\begin{theorem}\label{dyn}
\label{thm:dynamic} Consider the fixed-port model and let  $k(x)$ be
any function satisfying that $k(x)$, $\log_{k(x)} x$ and
$\frac{k(x)}{\log k(x)}$ are nondecreasing functions and that
$k(\Theta(x))=\Theta(k(x))$.\footnote{The above requirements are
satisfied by most natural sublinear functions such as $\alpha
x^{\epsilon}\log^{\beta} x$, $\alpha\log^{\beta}\log x$ etc..} There
exist dynamic 1-query labeling schemes supporting the distance,
separation level, id-NCA and routing functions on trees with the
following complexities.
\begin{enumerate}
\item
In the leaf-increasing tree model, with label size $O(\log_{k(n)}
n\cdot\log n)$ and amortized message complexity
$O(k(n)\cdot\log_{k(n)} n)$.
\item
In the leaf-increasing tree model, if an upper bound $n_f$ on the
number vertices in the dynamically growing tree is known in advance,
with label size $O(\frac{\log^3 n_f}{\log\log n_f})$  and amortized
message complexity $O(\frac{\log n_f}{\log\log n_f})$.
\item
In the leaf-dynamic tree model, with label size $O(\log_{k(n)}
n\cdot\log n)$ and amortized message complexity
$O\left(\sum_i{k(n_i)}\cdot\log_{k(n_i)} n\cdot
\frac{\MC(\pi,n_i)}{n_i}\right)+O(\sum_i\log^2 n_i)$.
\end{enumerate}
\end{theorem}

\section{Applications in a non-distributed environments}
In this section we first use ideas from our 1-query labeling schemes to
translate previously known results on finding NCA in trees to
support  distance queries. We then use the translated scheme to
show that general graphs can be
preprocessed efficiently to support flow queries.
\subsection{Distance queries in trees}
Harel and Tarjan \cite{HT84} describe a linear time algorithm to
preprocess a tree and build a data structure allowing NCA queries to
be answered in constant time on a RAM. Subsequently, simpler
algorithms with better constant factors have been proposed in
\cite{P90,BF00,GBT84,SV88,BV93}. On a pointer machine, \cite{HT84}
show a lower bound of $\Omega(\log\log n)$ on the query time, which
matches the upper bound of \cite{TL88}. In the leaf-increasing tree
model, \cite{G90,AT00} show how to make updates in amortized
constant time while keeping the constant worst-case query time on a
RAM, or the $O(\log\log n)$ worst-case query time on a pointer
machine. In \cite{CH05} they show how to maintain the above
mentioned results on a RAM in the leaf-dynamic tree model with worst
case constant update. See \cite{AGKR01} for a survey.

Simply by adding a pointer from each vertex to its depth and using
Equation 1, we obtain the following lemma.
\begin{lemma}
\label{non-distributed} The results of
\cite{HT84,P90,BF00,GBT84,SV88,BV93,HT84,TL88,G90,AT00,CH05} can
be translated to support either distance queries or separation
level queries (instead of NCA queries).
\end{lemma}
We note that other types of dynamic models were studied in the
non-distributed environment regarding NCA queries. See, for
example, \cite{CH05,AT00}. However, in these types of topological
changes, our transformation to distance queries is not efficient
since any such changes may effect the depth of too many vertices.

\subsection{Flow queries in general graphs}
Let $G\in\cG(n,W)$ and let $u_1,u_2,\cdots, u_n$ be the set of
vertices of $G$. Recall that in in \cite{KKKP04}, they show how to
construct a weighted tree $\tilde{T}_G\in\cT(O(n),W\cdot n)$ with
$n$ leaves $v_1,v_2,\cdots,v_n$ such that
$flow_G(u_i,u_j)=sep-level_{T_G}(v_i,v_j)$. Using Lemma
\ref{non-distributed} applied on the results of \cite{HT84}, we can
preprocess $\tilde{T}_G$ with $O(n\cdot\max\{1,\frac{\log W}{\log
n}\})$ space such that separation level queries can be answered in
$O(\max\{1,\frac{\log W}{\log n}\})$ time.
The exact model needed to prove the
following lemma formally is deferred to the full paper.
\begin{lemma}
Any graph $G\in\cG(n,W)$ can be preprocessed using
$O(n\cdot\max\{1,\frac{\log W}{\log n}\})$ space such that flow
queries can be answered in $O(\max\{1,\frac{\log W}{\log n}\})$
time.
\end{lemma}

\section{Conclusion and open problems}

In this paper we demonstrate that considering the labels of three
vertices, instead of two, can lead to a significant reduction in the
sizes of the labels. Inspecting two labels, and inspecting three,
are approaches that lie on one end of a spectrum. On the other end
of the spectrum would be a representation for which the decoder
inspects the labels of all the nodes before answering ($n$-query
labeling schemes). It is not hard to show that for any graph family
$\cF$ on $n$ node graphs, and for many functions (for example,
distance or adjacency),  one can construct an $n$-query labeling
scheme on $\cF$  using asymptotically optimal labels, i.e,
 $\log |\cF|/n + \Theta(\log n)$-bit labels (though the decoder may not be
 polynomial). The idea behind such a scheme is to enumerate the graphs in $\cF$
 arbitrarily. Then, given some $G\in\cF$ whose number in this
 enumeration is $i$, distribute the binary representation of  $i$
 among the vertices of $G$.
 In this way, given the labels of all nodes, the decoder can
reconstruct the graph and answer  the desired query. Therefore, a
natural question is to examine other points in this spectrum, i.e,
examine $c$-query labeling schemes for $1<c<n$.

There are other dimensions to the above question. For example, by
our definition, given the labels of two vertices, the $c$ vertices
that are chosen by the query algorithm $Q$ are chosen
simultaneously. Alternatively, one may define a possibly stronger
model in which these $c$ vertices are chosen one by one, i.e., the
next vertex is determined using the knowledge obtained from the
labels of previous vertices.

\newpage

\pagenumbering{roman} \setcounter{page}{1}
 \roman{page}

\bibliographystyle{plain}

\begin{thebibliography}{99}
\bibitem{AGM04}
I.~Abraham, C.~Gavoille, and D.~Malkhi.
 \newblock Routing with improved communication-space trade-off.
  \newblock {\em Oroc. 18th Int. Symp. on
Distributed Computing (DISC)}, Oct. 2004.


\bibitem{AACWY}
D.~Angluin, J.~Aspnes, J.~Chen, Y~Wu, Y.~Yin.
 \newblock Fast
construction of overlay networks.
 \newblock {\em Proc. SPAA 2005}, pp. 145-154.

\bibitem{AAPS96:jacm}
Y.~Afek, B.~Awerbuch, S.A.~Plotkin and M.~Saks.
\newblock Local management of a global resource in a communication.
\newblock {\em J. of the ACM }{\bf 43}, (1996), 1--19.

\bibitem{ABR03}
S.~Alstrup, P.~Bille and T.~Rauhe.
\newblock Labeling schemes for small distances in trees.
\newblock In {\em Proc. 14th ACM-SIAM Symp. on Discrete Algorithms},
  Jan. 2003.

\bibitem{AGKR01}
S.~Alstrup, C.~Gavoille, H.~Kaplan and T.~Rauhe.
\newblock Nearest Common Ancestors: A Survey and a new Distributed Algorithm.
\newblock {\em Theory of Computing Systems }{\bf 37}, (2004), 441--456.



\bibitem{AHT00}
S.~Alstrup, J.~Holm and M.~Thorup.
\newblock Maintaining Center and Median in Dynamic Trees.
\newblock In {\em Proc. 7th Scandinavian Workshop on Algorithm Theory},
  July. 2000.





\bibitem{AKM01}
S.~Abiteboul, H.~Kaplan and T.~Milo.
\newblock Compact labeling schemes for ancestor queries.
\newblock In {\em Proc. 12th ACM-SIAM Symp. on Discrete Algorithms},
  Jan. 2001.

\bibitem{AR02}
S.~Alstrup and T.~Rauhe.
\newblock Improved Labeling Scheme for Ancestor Queries.
\newblock In {\em Proc. 19th ACM-SIAM Symp. on Discrete Algorithms}, Jan. 2002.

\bibitem{AR02+}
S. Alstrup and T. Rauhe.
\newblock Small induced-universal graphs and compact implicit graph representations.
\newblock In {\em Proc. 43'rd annual IEEE Symp. on Foundations of Computer Science},
Nov. 2002.

\bibitem{AT00}
S.~Alstrup and M.~Thorup.
\newblock Optimal pointer algorithms for finding nearest common ancestors in dynamic trees.
\newblock {\em J. of Algorithms}, {\bf 35(2)}, (2000), 169--188.

\bibitem{Breuer66}
M.A.~Breuer.
\newblock Coding the vertexes of a graph.
\newblock {\em IEEE Trans. on Information Theory}, IT-12:148--153, 1966.

\bibitem{BF-67}
M.A.~Breuer and J.~Folkman.
\newblock An unexpected result on coding the vertices of a graph.
\newblock {\em J. of Mathematical Analysis and Applications }{\bf 20},
(1967), 583--600.


\bibitem{BF00}
M.A.~Bender and M.~Farach-Colton.
\newblock The LCA problem revised.
\newblock In {\em 4th LATIN}, 88--94, 2000.


\bibitem{BV93}
O.~Berkman and U.~Vishkin.
\newblock Recursive star-tree parallel data structure.
\newblock {\em SIAM J. on Computing }{\bf 22(2)},
(1993), 221--242.

\bibitem{C99}
L. J.~Cowen.
\newblock Compact Routing Schemes with Minimum Stretch.
\newblock In {\em Proc. 10th ACM-SIAM Symp. on Discrete Algorithms}, Jan. 1999.

\bibitem{CGK}
I.~Cidon, I.~S.~Gopal, and S.~Kutten.
 \newblock New models and algorithms
for future networks. \newblock {\em IEEE Transactions on
Information Theory} 41(3): 769-780 (1995).

\bibitem{CH05}
R.~Cole and R.~Hariharan.
\newblock Dynamic LCA Queries on Trees.
\newblock {\em SIAM J. on Computing }{\bf 34(4)}, (2005), 894-923.



\bibitem{CHKZ02}
E.~Cohen, E.~Halperin, H.~Kaplan and U.~Zwick.
\newblock Reachability and Distance Queries via 2-hop Labels.
\newblock In {\em Proc. 13th ACM-SIAM Symp. on Discrete Algorithms}, Jan. 2002.

\bibitem{CKM02}
E.~Cohen, H.~Kaplan and T.~Milo.
\newblock Labeling dynamic XML trees.
\newblock In {\em Proc. 21st ACM Symp. on Principles of Database
Systems}, June 2002.

\bibitem{EGI99}
D.~Eppstein, Z.~Galil and G. F.~Italiano.
\newblock Dynamic Graph Algorithms.
\newblock In {\em Algorithms and Theoretical Computing Handbook },
M.J. Atallah, Ed., CRC Press, 1999, Chapt. 8.

\bibitem{EGP96}
T.~Eilam C.~Gavoille and D.~Peleg.
\newblock Compact Routing Schemes with Low Stretch Factor.
\newblock In {\em Proc. 17th Annual ACM Symp. on Principles of Distributed Computing },
ACM Press, may 1996, 11--20.


\bibitem{G90}
H. N.~Gabow.
\newblock Data Structure for Weighted Matching and Nearest Common
Ancestor with Linking.
\newblock In {\em Proc. 1st Annual ACM Symp. on Discrete Algorithms }, Jan. 1990, 434--443.

\bibitem{G00}
C.~Gavoille.
\newblock A Survey on Interval Routing.
\newblock In {\em Theoretical Computer Science },
{\bf 245}, (2000), 217--253.



\bibitem{FG01}
P.~Fraigniaud and C.~Gavoille.
\newblock Routing in trees.
\newblock In {\em Proc. 28th Int. Colloq. on Automata, Languages \&
Prog.}, LNCS 2076, pages 757--772, July 2001.

\bibitem{FG02}
P.~Fraigniaud and C.~Gavoille.
\newblock  A space lower bound for routing in trees.
\newblock In {\em Proc. 19th Int. Symp. on Theoretical Aspects of Computer
Science}, March 2002, 65-75.

\bibitem{debrujin-pierre}
P.~Fraigniaud and P.~Gauron.
 \newblock  D2B: A de Bruijn based
content-addressable network.
 \newblock Theor. Comput. Sci. 355(1): 65-79
(2006).

\bibitem{FK00}
J.~Feigenbaum and S.~Kannan.
\newblock Dynamic Graph Algorithms.
\newblock In {\em Handbook of Discrete and Combinatorial Mathematics}, CRC Press, 2000.


\bibitem{GBT84}
H. N.~Gabow, J. L.~Bentley and R. E.~Tarjan.
\newblock Scaling and related techniques for geometry problems.
\newblock In {\em Proc.  16th Annual ACM Symp. on Theory of Computing}, May 1984.



\bibitem{GP01a}
C.~Gavoille and C.~Paul.
\newblock Split decomposition and distance labeling: an optimal
scheme for distance hereditary graphs.
\newblock In {\em Proc.  European Conf. on Combinatorics, Graph Theory and
  Applications}, Sept. 2001.

\bibitem{GP01b}
C.~Gavoille and D.~Peleg.
\newblock Compact and Localized Distributed Data Structures.
\newblock {\em J. of Distributed Computing }{\bf 16}, (2003), 111--120.

\bibitem{GKKPP01}
C.~Gavoille, M.~Katz, N.A.~Katz, C.~Paul and D.~Peleg.
\newblock Approximate Distance Labeling Schemes.
\newblock In {\em 9th European Symp. on Algorithms}, Aug. 2001,
Aarhus, Denmark, SV-LNCS 2161, 476--488.

\bibitem{GPPR01}
C.~Gavoille, D.~Peleg, S.~P\'erennes and R.~Raz.
\newblock Distance labeling in graphs.
\newblock  {\em J. of Algorithms} {\bf 53(1)} (2004), 85--112.

\bibitem{H80}
H.~Harel.
\newblock A linear time algorithm for lowest common ancestors problem.
\newblock In {\em 21st Annual IEEE  Symp. on Foundation of Computer Science}, Nov. 1980.

\bibitem{HLL}
M.~Harchol-Balter, F.~T.~Leighton, and D.~Lewin.
 \newblock Resource
Discovery in Distributed Networks.
 \newblock {\em Proc. PODC} 1999, pp. 229-237.

\bibitem{HLT01}
J.~Holm, K.~Lichtenberg and M.~Thorup.
\newblock Poly-logarithmic deterministic fully-dynamic algorithms for connectivity, minimum spanning tree,
2-edge, and biconnectivity.
\newblock {\em J. of the ACM }{\bf 48(4)}, (2001), 723--760.

\bibitem{HT84}
H.~Harel and R. E.~Tarjan.
\newblock Fast algorithms for finding nearest common ancestors.
\newblock {\em SIAM J. on Computing}, {\bf 13(2)}, 1984, 338--355.

\bibitem{subpages}
H.~A.~Jamrozik, M.~J.~Feeley, G.~M.~Voelker, J.~Evans, A.~R.~Karlin,
H.~M.~Levy, and M.~K.~Vernon.
\newblock Reducing network latency using subpages
in a global memory environment.
\newblock {\em Proc. the 7th ACM
Conference on Architectural Support for Programming Languages and
Operating Systems}, 1996.


\bibitem{K05}
A.~Korman.
\newblock General Compact Labeling Schemes for Dynamic Trees.
\newblock In {\em Proc. 19th International Symp. on Distributed Computing},
  Sep. 2005.

\bibitem{KKKP04}
M.~Katz, N.A.~Katz, A.~Korman and D.~Peleg.
\newblock Labeling schemes for flow and connectivity.
\newblock {\em SIAM Journal on Computing} {\bf 34} (2004),23--40.

\bibitem{KKP00}
M.~Katz, N.A.~Katz, and D.~Peleg.
\newblock Distance labeling schemes for well-separated graph classes.
\newblock In {\em Proc. 17th Symp. on Theoretical Aspects of Computer Science},
  pages 516--528, Feb. 2000.

\bibitem{KP03}
A.~Korman and D.~Peleg.
\newblock Labeling Schemes for Weighted Dynamic Trees.
\newblock In {\em Proc. 30th Int. Colloq. on Automata, Languages \& Prog.},
Eindhoven, The Netherlands, July 2003, SV LNCS.

\bibitem{KPR04}
A.~Korman, D.~Peleg and Y.~Rodeh.
\newblock Labeling schemes for dynamic tree networks.
\newblock {\em Theory of Computing Systems }{\bf 37}, (2004), 49--75.



\bibitem{KNR92}
S.~Kannan, M.~Naor, and S.~Rudich.
\newblock Implicit representation of graphs.
\newblock In {\em SIAM J. on Discrete Math }{\bf 5}, (1992), 596--603.

\bibitem{KM01}
H.~Kaplan and T.~Milo.
\newblock Short and simple labels for small distances and other functions.
\newblock In {\em Workshop on Algorithms and Data Structures}, Aug. 2001.

\bibitem{KM01a}
H.~Kaplan and T.~Milo.
\newblock Parent and ancestor queries using a compact index.
\newblock In {\em Proc. 20th ACM Symp. on Principles of Database Systems},
  May 2001.

\bibitem{KMS02}
H.~Kaplan, T.~Milo and R.~Shabo.
\newblock A Comparison of Labeling Schemes for Ancestor Queries.
\newblock In {\em Proc. 19th ACM-SIAM Symp. on Discrete Algorithms}, Jan. 2002.


\bibitem{ercim}
E.~P. Markatos and G.~Dramitinos.
\newblock
Remote Memory to Avoid Disk Thrashing: A
Simulation Study.
\newblock {\em Proc. MASCOTS} 1996: 69-73.


\bibitem{viceroy}
D.~Malkhi, M.~Naor, and D.~Ratajczak.
 \newblock Viceroy: A scalable and dynamic emulation of the
 butterfly.
  \newblock {\em Proc. 21st annual ACM symposium
    on Principles of distributed computing}, 2002.


\bibitem{debrujin-udi}
 Moni Naor and Udi Wieder.
  \newblock Novel architectures for P2P
applications: the continuous-discrete approach.
 \newblock {\em Proc. SPAA} 2003, pp. 50-59.


\bibitem{P99:lbl}
D.~Peleg.
\newblock Proximity-preserving labeling schemes and their applications.
\newblock In {\em Proc. 25th Int. Workshop on Graph-Theoretic Concepts in
  Computer Science}, pages 30--41, June 1999.

\bibitem{Peleg00:lca}
D.~Peleg.
\newblock Informative labeling schemes for graphs.
\newblock In {\em Proc. 25th Symp. on Mathematical Foundations of Computer
  Science}, volume LNCS-1893, pages 579--588. SV, Aug. 2000.

\bibitem{Peleg00:book}
D.~Peleg.
\newblock {\em Distributed Computing: A Locality-Sensitive Approach}.
\newblock SIAM, 2000.

\bibitem{P90}
P.~Powell.
\newblock A further improved LCA algorithm.
\newblock Technical Report TR90-01, University of Minneapolis, 1990.

\bibitem{CAN}
S.~Ratnasamy, P.~Francis, M.~Handley, R.~Karp, and S.~Shenker.
\newblock A scalable content-addressable network.
\newblock {\em Proc. ACM SIGCOMM}
2001, pp. 161-172, August 2001.


\bibitem{SK85}
N.~Santoro and R. Khatib.
\newblock Labeling and implicit routing in networks.
\newblock {\em The Computer Journal }{\bf 28}, (1985), 5--8.

\bibitem{Chord} I.~Stocia, R.~Morris, D.~Karger, F.~Kaashoek, and
H.~Balakrishnan.
 \newblock Chord: A Scalable Peer-to-peer Lookup
Service for Internet Applications.
 \newblock {\em Proc. ACM SIGCOMM}
2001, San Diego, CA, Aug. 2001.


\bibitem{ST83}
D. D.~Sleator and R. E.~Tarjan.
\newblock A data structure for dynamic trees.
\newblock {\em Journal of Computer and System Sciences }{\bf 26(1)}, (1983), 362--391.

\bibitem{SV88}
B.~Schieber and U.~Vishkin.
\newblock On finding lowest common ancestors: Simplification and parallelization.
\newblock {\em SIAM J. on Computing}, {\bf 17}, 1988, 1253--1262.



\bibitem{T01}
M.~Thorup.
\newblock Compact oracles for reachability and approximate distances
in planar digraphs.
\newblock {\em J. of the ACM }{\bf 51}, (2004), 993--1024.

\bibitem{TL88}
A. K. ~Tsakalides and J. van Leeuwen.
\newblock An optimal pointer machine algorithm for finding nearest common ancestors.
\newblock Technical Report RUU-CS-88-17, Department of CS, University of Utrecht, 1988.



\bibitem{TZ01}
M.~Thorup and U.~Zwick.
\newblock Compact routing schemes.
\newblock In {\em Proc. 13th ACM Symp. on Parallel Algorithms and
  Architecture}, pages 1--10, Hersonissos, Crete, Greece, July 2001.

\bibitem{VT87}
J.~Van Leeuwen and R. B.~Tan.
\newblock Interval routing.
\newblock {\em The Computer Journal }{\bf 30}, (1987), 298--307.

\bibitem{tapestry}
B.~Zhao, J.~Kubiatowicz,  and A.~Joseph.
 \newblock Tapestry: An
Infrastructure for Fault-tolerant Wide-area Location and Routing.
 \newblock Tech. rep., University of California, Berkeley, 2001.

\end{thebibliography}

\appendix
\section{Appendix}

\subsection{Definition of the functions}

We consider the following functions which are applied on pairs of
vertices $u$ and $v$ in a graph $G=\langle V,E\rangle$. (When the
underlying graph is clear from the context we omit the corresponding
subscript).
\begin{itemize}

\item{Flow:} Denote by $G'$ the multigraph obtained by replacing
each edge $e$ in $G$ with $\omega(e)$ parallel edges of capacity
1. A set of paths $P$ in $G'$ is {\em edge-disjoint} if each edge
$e\in E$ appears in no more than one path $p \in P$. Let
$\cP_{u,v}$ be the collection of all sets $P$ of edge-disjoint
paths in $G'$ between $u$ and $v$. Then the maximum flow between
$u$ and $v$ is defined as $f(u,v) ~=~ \max_{P \in
\cP_{u,v}}\{|P|\}$, where $|P|$ is the number of paths in $P$. See
\cite{KKKP04} for additional details regarding labeling for the
flow function.

\item{distance:} in the case where $G$ is weighted
(respectively, unweighted), $d_G(u,v)$ equals the weighted (resp.
unweighted) distance between $u$ and $v$ in $G$.

\item{routing:}
function: $rout_G(u,v)$ is the port number at $u$ leading to the
next vertex on the shortest path
connecting $u$ to $v$.\\
\end{itemize}

\noindent
 If the graph is a tree $T$ then we consider also the
following functions:

\begin{itemize}
\item{separation level:}$Sep-level_T(u,v)$ equals the depth of $NCA(u,v)$.
\end{itemize}

\noindent
 We distinguish between the following two variants for the
NCA function.

\begin{itemize}

\item{id-NCA:} assume that identities containing
$O(\log n)$ bits are assigned to the vertices by an adversary. Then,
$id-NCA_T(u,v)$ is the identity of $NCA(u,v)$.

\item{label-NCA:}  assuming each vertex can freely
select its own identity (as long as all identities remain unique),
$label-NCA_T(u,v)$ is the identity of $NCA(u,v)$. (In this case, the
identities may also be referred to as labels.)
\end{itemize}

\subsection{Query labeling schemes for the separation level and
distance functions} We start with describing a the 1-query labeling
scheme
$\varphi_{sep-level}=\langle\cM_{sep-level},Q_{sep-level},\cD_{sep-level}\rangle$
which supports the separation level function on $\cT(n,W)$. The idea
behind $\varphi_{sep-level}$ is to label the first sublabels with
the labels given by the label-NCA labeling scheme $\pi_{NCA}(v)$ and
to label the second sublabel of each vertex $v$ with its depth
$depth(v)$. Then, using the labels of two vertices $u$ and $v$, the
query algorithm $Q_{sep-level}$ outputs the first sublabel of
$w=NCA(u,v)$ using the decoder of $\pi_{NCA}(v)$. Given the labels
of $u$, $v$ and $w$, the decoder $\cD_{sep-level}$ simply outputs
the second sublabel of $w$, which is its depth. Since the label size
of $\pi_{NCA}(v)$ is $O(\log n)$ and since the depth of each vertex
$v$ can be encoded using $O(\log (nW))$ bits, we obtain that the
label size of $\varphi_{sep-level}$ is $O(\log n+\log W)$. Using
Claim \ref{lower-bound}, we obtain the following claim.

\begin{claim}\label{sep}
The label size of a 1-query separation-level labeling scheme on
$\cT(n,W)$ is $\Theta(\log n+\log W)$.
\end{claim}

We now describe how to transform our 1-query labeling scheme
$\varphi_{sep-level}$ into a 1-query distance labeling scheme
$\varphi_{dist}=\langle\cM_{dist},Q_{dist},\cD_{dist}\rangle$. The
marker and query algorithms of $\varphi_{dist}$ are the same as the
corresponding marker and query algorithms of $\varphi_{sep-level}$.
Given the labels $\cM_{dist}(v)$, $\cM_{dist}(u)$ of two vertices
$v$ and $u$ in some $T\in\cT(n,W)$ as well as the label
$\cM_{dist}(w)$ of $w=NCA(v,u)$, the decoder algorithm $\cD_{dist}$
outputs
$\cM_{dist}^{data}(v)+\cM_{dist}^{data}(u)-2\cM_{dist}^{data}(w)$.
The correctness of $\varphi_{dist}$ follows from the following
equation.
\begin{equation}
d(v,u)=depth(v)+depth(u)-2\cdot depth(NCA(v,u)).
\end{equation}
 Since the
marker algorithm of $\varphi_{dist}$ is the same as the marker
algorithm of $\varphi_{sep-level}$, we obtain that the label size of
$\varphi_{dist}$ is $O(\log n+\log W)$. The following lemma follows,
using Claim \ref{lower-bound}.
\begin{claim}\label{dist}
The label size of a 1-query distance labeling scheme on $\cT(n,W)$
is $\Theta(\log n+\log W)$.
\end{claim}

Lemma \ref{sep+dist} follows by combining Claim \ref{sep} and Claim
\ref{dist}.
 \subsection{Sketch of the proof of Lemma
\ref{lem:1qflow}} The fact that any 1-query flow labeling scheme on
$\cG(n,W)$ must have label size $\Omega(\log n+\log W)$ follows
using similar arguments as in the proof of Claim \ref{lower-bound}.
We now show how to construct a 1-query flow labeling scheme on
$\cG(n,W)$ with label size $O(\log n+\log W)$. As shown in
\cite{KKKP04}, given a graph $G\in\cG(n,W)$ with vertices
$u_1,u_2,\cdots, u_n$, one can construct a weighted tree
$\tilde{T}_G\in\cT(O(n),W\cdot n)$ with $n$ leaves
$v_1,v_2,\cdots,v_n$ such that
$flow_G(u_i,u_j)=sep-level_{T_G}(v_i,v_j)$. Therefore, using our
1-query separation level labeling scheme $\varphi_{sep-level}$ on
$\tilde{T}_G$, we obtain a  1-query flow labeling scheme
$\varphi_{flow}$ with size $O(\log(O(n))+\log (nW))=O(\log n+\log
W)$.\QED

\subsection{A 3-approximation routing scheme in the fixed port
model} First note that in order to make sense, any routing scheme
must have label size  $\Omega(\log n)$ since the labels must be
distinct. Moreover, the header size of any routing scheme must also
be $\Omega(\log n)$ since the label of the destination vertex needs
to be encoded in the header. We now construct our 3-approximation
routing labeling scheme $\pi_{approx}=\langle
\cM_{approx},\cR_{approx}\rangle$, which operates in the fixed-port
model and has label size and header size $\Theta(\log n)$.

Recall that the designer port numbers at any vertex $v$ are
numbered from 1 to $deg(v)$ (except that the  designer port
numbers at the root $r$ are numbered from  2 to $deg(r)$).
Moreover, the designer port number leading from any non-root
vertex $v$ to its parent is numbered 1. Let $v\in T$ be a non-leaf
vertex. For every $2\leq i\leq deg(v)$, let $child^{des}_i(v)$ be
the child of $v$ such that the designer port number at $v$ leading
to $child^{des}_i(v)$ is $i$, i.e.,
$port_{des}(v,child^{des}_i(v))=i$. Let $c(v)$ denote the number
of children of $v$, i.e, $c(v)=deg(v)$ if $v$ is the root and
$c(v)=deg(v)-1$ otherwise. For every $1\leq i\leq c(v)$, let
$port^{fix}_i(v)$ be the $i$'th smallest fixed-port number at $v$
among the fixed-port numbers leading from $v$ to its children. For
every $1\leq i\leq c(v)$, let $child^{fix}_i(v)$ be the child of
$v$ such that $port^{fix}_i(v)$ is the fixed-port number at $v$
leading to $child^{fix}_i(v)$, i.e.,
$port_{fix}(v,child^{fix}_i(v))=port^{fix}_i(v)$. Let $n(v)$
denote the number $i$ such that $v=child^{fix}_i(p(v))$. The
marker algorithm $\cM$ assigns $v$ the label
$$\cM_{fix}(v)=\langle\cM_{des}(v)~,~
port_{fix}(p(v)~,~child^{des}_{n(v)}(p(v)))~,~
port_{fix}(v,p(v))\rangle.$$

The label of the root $r$ is simply
$\cM_{fix}(r)=\langle\cM_{des}(r),0,0\rangle$. Note that the label
of a  vertex $v$ is composed of three fields, namely,
$\cM_{fix}(v)=\langle\cM_1(v),\cM_2(v),\cM_3(v)\rangle$. For a
non-root vertex $v$, its second field $\cM_2(v)$ is the fixed-port
number of the port at $v$'parent whose designer port number is
$n(v)$. (Recall that $n(v)$ was defined to be the {\em fixed} port
leading to $v$.)

The routing is performed as follows. Given the label $\cM(w)$ of the
destination vertex $w$ and the label $\cM(v)$ of the vertex $v$
initiating the route of the message, the router algorithm $\cR$ at
$v$ does the following. Let $i=\cD_{des}(\cM_1(v),\cM_1(w))$. If
$i=1$ then a header containing $\langle \cM(w), 0\rangle$ is
attached to the message and the message (with the header) is
forwarded to $v$'s parent via port number $\cM_3(v)$. If $i\neq 1$
then  a header containing $\langle \cM(w),-1\rangle$ is attached to
the message, and the message (with the header) is forwarded to $v$'s
child $child^{fix}_i(v)$ via the $i$'th smallest adversary port
number at $v$ among the fixed-port numbers leading to children of
$v$. Note that since $\cM_3(v)$ is the fixed-port number leading
from $v$ to its parent, $v$ can identify which of its ports lead to
its children. Furthermore, it is a local computation at $v$ to find
which of its (adversary) port numbers is the $i$'th smallest port
number among the  port numbers leading to its children.

If a vertex $v$ receives a message with header $\langle \cM(w),
0\rangle$ and $\cM(v)=\cM(w)$ then $v=w$ and the message reached its
destination $w$ thus completing the route. If, $\cM(v)\neq\cM(w)$
then the router $\cR$ at $v$ does the following (similar operations
to the case $v$ is the vertex initiating the route). Let
$i=\cD_{des}(\cM_1(v),\cM_1(w))$. If $i=1$ then a header is attached
to the message containing $\langle \cM(w), 0\rangle$ and the message
(with the header) is forwarded to $v$'s parent via port number
$\cM_3(v)$. If $i\neq 1$ then  a header is attached to the message
containing $\langle \cM(w), -1\rangle$ and the message (with the
header) is forwarded to $v$'s child $child^{fix}_i(v)$ via port
number $port_i^{fix}(v)$.

If a vertex $v$ receives a message with header $\langle \cM(w),
-1\rangle$ then the router $\cR$ at $v$ does the following. A header
containing $\langle \cM(w), \cM_2(v)\rangle$ is attached to the
message and the message (with the header) is forwarded to $v$'s
parent via port number $\cM_3(v)$.

If a vertex $v$ receives a message with header $\langle \cM(w),
x\rangle$ , where $x\neq 0,-1$, then the router $\cR$ at $v$ does
the following. A header  containing $\langle \cM(w), 0\rangle$ is
attached to the message and the message (with the header) is
delivered to a child of $v$  via port number $x$. Let us now show
correctness.
\begin{lemma}
The routing  scheme $\pi_{approx}=\langle
\cM_{approx},\cR_{approx}\rangle$ is a $3$-approximation routing
scheme.
\end{lemma}
\begin{proof}
Let $w$ be the destination vertex. Any intermediate node $u$ on the
route delivers the message to its parent iff its parent is the next
vertex on the shortest path connecting $u$ and $w$. Furthermore, if
a vertex $u$ delivers the message to its parent then the header of
the message is set to $\langle \cM(w),0\rangle$.

Any intermediate node $u$ receiving the message with header
$\langle\cM(w),x\rangle$, where $x\neq 0,-1$, deliverers the message
to  one of its children via port number $x$, where $x$ is precisely
the fixed-port number of the port at $v$ whose designer port number
is $\cD_{des}(\cM_{des}(v),\cM_{des}(w))$. Therefore, the message is
delivered to the next vertex on the shortest path connecting $u$ and
$w$. Moreover, the message is delivered with the header $\langle
\cM(w),0\rangle$.

 If $u$ is the vertex initiating the route or if $u$ is some
intermediate node receiving the message with header $\langle
\cM(w),0\rangle$, where the next vertex on the shortest path from
$u$ to $w$ is one of $u$'s children, then precisely two messages are
sent before the message returns to $u$ with header
$\langle\cM(w),x\rangle$, where $x\neq 0,-1$.

We therefore get that the resulting route from the initiator $v$ to
the destination $w$ is at most thrice as long as the shortest path
connecting $v$ and $w$.\QED
\end{proof}
Note that the label size of $\pi_{des}$ is $O(\log n)$ and that each
port number is  encoded using  $O(\log n)$ bits. Therefore, using
the previous lemma, we obtain Lemma \ref{3-approx}

\subsection{Sketch of the Proof of Theorem \ref{dyn}} The proof
presented here in not completely self contained. A truely complete
proof would include many details concerning three papers, namely,
\cite{FG01} and especially \cite{KPR04} and \cite{K05}, and is,
thus, too long to be given here.

We divide the proof into two parts. In the first part we consider
the claims of Theorem \ref{dyn} regarding the dynamic 1-query
labeling schemes supporting the distance, separation level and
id-NCA functions and in the second part we consider the claims of
Theorem \ref{dyn} regarding the dynamic 1-query routing labeling
scheme. \\
\subsubsection{First part of the proof} Let us first consider the
claims of Theorem \ref{dyn} regarding the dynamic 1-query labeling
schemes supporting the distance, separation level and id-NCA
functions. This part follows rather naturally from
\cite{KPR04,K05}.

Recall that in a query labeling scheme, the label of each vertex is
composed of two sublabels, the index sublabel and the data sublabel.
In our (static) 1-query labeling schemes supporting the distance,
separation level and id-NCA functions, the index sublabel of each
vertex $v$ is simply the label assigned to $v$ by ${\pi}_{NCA}$, the
label-NCA labeling scheme of \cite{AGKR01}, and the data sublabel of
$v$ is either $v$'s depth or $v$'s identity. Therefore, we can
dynamically maintain the index sublabel of each vertex using
$\hat{\pi}_{NCA}$, the dynamic label-NCA labeling scheme resulted by
applying either Theorem 4.16 of \cite{KPR04} or Theorems 1 and 2 of
\cite{K05} on ${\pi}_{NCA}$. In addition, the data sublabel of each
vertex $v$ can easily be maintained as follows. In case we consider
the id-NCA function, whenever a new vertex $v$ joins the tree and an
identity $id(v)$ is given to $v$ by the adversary, this identity is
stored at the data sublabel of $v$. If we consider, instead,
 either the distance or the separation level functions,
the data sublabel of the root is set to be 0, and whenever a new
vertex $v$ joins the tree, it communicates with its parent $p(v)$
and sets its data sublabel to be
$\hat{\cM}_{data}(v)=1+\hat{\cM}_{data}(p(v))$, where
$\hat{\cM}_{data}(p(v))$ is the data sublabel given to $p(v)$ by our
dynamic 1-query scheme. Therefore, each vertex $v$ maintains its
depth (or identity) in the its data sublabel with an extra constant
additive cost to the amortized message complexity of
$\hat{\pi}_{NCA}$.

The query and decoder algorithms of the resulted dynamic 1-query
labeling scheme relate to $\hat{\pi}_{NCA}$ similarly to the way
the query and decoder algorithms of the corresponding static
1-query scheme relate to ${\pi}_{NCA}$. Since the labels of
$\pi_{NCA}$ can be assigned by a distributed algorithm using
$O(n)$ messages, Theorem 4.16 of \cite{KPR04} and Theorems 1 and 2
of \cite{K05} imply the claims of Theorem \ref{dyn} regarding the
dynamic 1-query labeling schemes supporting the
distance, separation level and id-NCA functions.\\
\subsubsection{Second part of the proof} We now consider the
claims of Theorem \ref{dyn} regarding the dynamic 1-query routing
labeling scheme. Let us first recall the scheme  $\pi_{des}$ of
\cite{FG01} for routing over trees, in which each vertex has both a
routing table and  a  label. When routing a message from vertex $x$
to a vertex $y$, vertex $x$ is given the label $L(y)$ of the
destination vertex $y$ and uses its routing table to decide which of
its incident ports leads to the next vertex $w$ on the shortest path
connecting $x$ and $y$. Then $x$ prepares a header containing the
label $L(y)$ and sends the message together with the header to $w$.
When $w$ receives the message it repeats this process (without
changing the header) until the message reaches $y$. Since both the
routing table and the label of each vertex use $O(\log n)$ bits, one
can encode the routing table into the label and easily modify
$\pi_{des}$ into a routing labeling scheme with label size $O(\log
n)$. Using the terminology of \cite{FG01}, the label of each vertex
$v$ is $\langle id(v),\omega(v),\omega_1(v),cpath(v)\rangle$. In
particular, the label of the root $r$ is $\langle
1,n,\omega_1(r),\emptyset\rangle$, where $\omega_1(r)$ is the
largest number of descendants of any child of $r$.

We now describe how to modify $\pi_{des}$ to be a routing labeling
scheme on $n$-vertices trees with label size $O(\log n)$, such that
the label of the root is always $\langle 1\rangle$. We denote the
modified scheme by $\pi'_{des}$. We note that our dynamic query
labeling scheme (described label) will invoke $\pi'_{des}$ on
multiple subtrees of the given dynamic tree. In each such
application, the label given to the root of the subtree (not just
the root of the whole tree) is always $\langle 1\rangle$ (this label
is then concatenated  with an additional label, to obtain unique
labeling). For every vertex $v$, let $L(v)$ denote the label given
to $v$ by $\pi_{des}$ and for every non-root vertex $v$, let
$\rho(v)$ be the port number leading from the root $r$ to the next
vertex on the shortest path connecting $r$ and $v$. The label given
to any vertex $v$ by $\pi_{des}'$ is the following:
$$
L'(v) ~\gets~ \left\{
\begin{array}{ll}
1 \qquad & \mbox{if } v = r, \\
L(v)\circ \rho(v) \qquad & \mbox{otherwise, where }\circ\mbox{ stands for concatenation.}  \\
\end{array}
\right.
$$
Let us mention that in $\pi_{des}$, the identity $id(v)$ of the a
vertex $v$ is its DFS number. Hence, $L'(r)=1=id(r)$. As shown in
\cite{FG01},  given the labels $L(x)$ and $L(y)$ of two vertices $x$
and $y$ in the tree, the decoder $\cD$ of $\pi_{des}$ operates as
follows.

$$
\cD(L(x),L(y)) ~\gets~ \left\{
\begin{array}{ll}
0 \qquad & \mbox{if } id(y)=id(x), \\
1 \qquad & \mbox{if } id(y)<id(x)\mbox{ or } id(y)\geq id(x)+w(x),\\
1+b \qquad & \mbox{if } id(x)<id(y)\leq id(x)+w_1(x), \\
p+b \qquad & \mbox{otherwise, where } p=|cpath(x)+1|\mbox{-th element of } cpath(y).\\
\end{array}
\right.
$$

Without getting into the details concerning the meanings of the
parameters in the above formula, we describe the decoder $\cD'$ of
$\pi_{des}'$ which satisfies $\cD'(L'(x),L'(y))=\cD(L(x),L(y))$ for
every two vertices $x$ and $y$. Given the labels $L'(x)$ and $L'(y)$
of two vertices $x$ and $y$, the decoder $\cD'$ operates as follows.

$$
\cD'(L'(x),L'(y)) ~\gets~ \left\{
\begin{array}{ll}
0 \qquad & \mbox{if } id(y)=id(x), \\
\cD(L(x),L(y))\qquad & \mbox{if } L'(x)\neq \langle 1\rangle\mbox{ and }L'(y)\neq \langle 1\rangle, \\
1 \qquad & \mbox{if } id(y)=\langle 1\rangle \mbox{ and } id(x)>1,\\
\rho(y) \qquad & \mbox{if } id(x)=\langle 1\rangle \mbox{ and } id(y)>1. \\
\end{array}
\right.
$$

The following claim shows that $\pi_{des}'$ is a correct
routing labeling scheme on trees.\\
{\bf Claim 1:} For every two
vertices $x$ and $y$, $\cD'(L'(x),L'(y))=\cD(L(x),L(y))$.\\
{\bf Proof of Claim 1:} Clearly, if $id(x)=id(y)$ then
$\cD'(L'(x),L'(y))=\cD(L(x),L(y))=0$. Moreover, if both $x\neq r$
and $y\neq r$, then by the definition of the decoder $D'$,
$\cD'(L'(x),L'(y))=\cD(L(x),L(y))$. Assume now that $id(x)=1$
($x=r$) and $id(y)>1$ ($y\neq r$). In this case, by the definition
of $\rho(y)$, $D'(L'(x),L'(y))=\rho(y)$ is the port number leading
from $x=r$ to the next vertex on the shortest path connecting $x$
and $y$. By the correctness of $\pi_{des}$, $D(L(x),L(y))$ also
equals $\rho(y)$. If $id(y)=1$  and $id(x)>1$, then $y$ is the root
and therefore the parent of $x$ is next vertex on the shortest path
connecting $x$ and $y$. It follows that $D(L(x),L(y))=1$, and by the
definition of the decoder $\cD'$, $D'(L'(x),L'(y))$ also equals 1.
This established Claim 1.\QED

Since the label size of $\pi_{des}$ is $O(\log n)$ and since for
every vertex $v$, the port number $\rho(v)$ can be encoded using
$O(\log n)$ bits, we obtain the following claim.\\ {\bf Claim 2:}
$\pi_{des}'$ is a routing labeling scheme on $\cT(n)$ with label
size $O(\log n)$. Moreover, the label given to the root of any tree
by the scheme $\pi_{des}'$ is $\langle 1\rangle$.

We now describe how to  extend $\varphi_{fix}$, our 1-query
routing labeling scheme, to the dynamic scenario. We denote the
resulted scheme by $\hat{\varphi}_{fix}=\langle
\hat{\cM}_{fix},\hat{Q}_{fix},\hat{\cD}_{fix}\rangle$. Let $T$ be
a dynamic tree in the fixed-port model. Let $\hat{\pi}_{des}(v)$
be the dynamic routing labeling scheme resulted by applying either
Theorem 4.16 of \cite{KPR04} or Theorems 1 and 2 of \cite{K05} on
the modified routing labeling scheme $\pi_{des}'$ described above.
Note that $\pi_{des}'$ is designed to operate in the designer port
model, and therefore $\hat{\pi}_{des}=\langle
\hat{\cM}_{des},\hat{\cD}_{des}\rangle$ should also operate in the
designer port model. However, our aim is a scheme for the fixed
port model. Informally, we overcome this difficulty by running
$\hat{\pi}_{des}$ assuming the designer port model and whenever
the marker algorithm $\hat{\cM}_{des}$ wishes to assign a port
number to a port of $v$ (as it should in the designer port model),
it instead uses this value to label the corresponding child of
$v$. Formally, the dynamic marker algorithm $\hat{\cM}_{fix}$
operates as follows. As in any query labeling scheme, the label
$\hat{\cM}_{fix}(v)$ assigned to a vertex $v$ by the dynamic
marker algorithm $\hat{\cM}_{fix}$ contains two sublabels, namely,
the index sublabel $\hat{\cM}^{index}(v)$ and the data sublabel
$\hat{\cM}^{data}(v)$. For simplicity, we assume that the dynamic
tree contains only the root when the algorithm starts. The case
that the algorithm is started when the tree already contains
additional vertices is deferred to the full paper. (It uses some
additional modification of the translation methods.)

\subsubsection*{The dynamic marker algorithm $\hat{\cM}_{fix}$}
The dynamic marker algorithm $\hat{\cM}_{fix}$ first invokes
$\hat{\cM}_{des}$ as if we where in the designer port model. We do
not describe here the dynamic algorithm $\hat{\cM}_{des}$ (given
either by the method of \cite{KPR04} or by the method of
\cite{K05}). However, we note that from time to time
$\hat{\cM}_{des}$ assigns and updates labels and port numbers as a
result of a recomputation (shuffle- in the terminology of
\cite{KPR04} and reset- in the terminology of \cite{K05}) performed
on some subtree. The events of assigning and updating the labels and
port numbers are modified in the dynamic marker algorithm
$\hat{\cM}_{fix}$ as described in the following steps (which are
applied simultaneously). In particular, the port numbers that should
be assigned by the marker algorithm $\hat{\cM}_{des}$, are not
assigned to the ports (since we are dealing with the fixed-port
model). Instead, these numbers are used as described in Steps 4 and
5 below.
\begin{enumerate}
\item Whenever a label $\hat{\cM}_{des}(v)$ is assigned to (or
updated at) a vertex $v$ by $\hat{\cM}_{des}$, this label is
stored at the first field of $v$'s data sublabel, i.e.,
$\hat{\cM}^{data}_1(v)=\hat{\cM}_{des}(v)$. \item Each time a
non-leaf vertex $v$ is supposed to be assigned a new label
$\hat{\cM}_{des}(v)$ under the dynamic marker algorithm
$\hat{\cM}_{des}$, it sends a message to each of its children
containing this new value $\hat{\cM}_{des}(v)$. In turn, each
child $u$ of $v$ sets the first field in its index sublabel to be
this value, i.e.,
$\hat{\cM}^{index}_1(u)=\hat{\cM}_{des}(v)=\hat{\cM}_{des}(p(u))$.
\item Whenever a new leaf $u$ joins the tree, it sends a signal to
its parent $v$ which in turn sends $u$ a message containing
$\hat{\cM}^{data}_1(v)$. When $u$ receives this message it sets
the first field in its index sublabel to be the value contained in
the message, i.e.,
$\hat{\cM}^{index}_1(u)=\hat{\cM}^{data}_1(v)=\hat{\cM}_{des}(p(u))$.
 \item Whenever the marker algorithm $\hat{\cM}_{des}$ wishes to
assign a port number $\rho$ to a port leading from a vertex $v$ to
one of its children $u$, it refrains from doing so and instead
assigns the second field of $\hat{\cM}^{index}(u)$ the value
$\rho$, i.e., $\hat{\cM}^{index}_2(u)=port_{des}(p(u),u)$. \item
Whenever a leaf $u$ joins the tree as a child of some vertex $v$
and the corresponding  ports are assigned a port number by the
adversary, the following happens. The port number
$port_{fix}(u,v)$ is stored at the third field of the data
sublabel of $u$, i.e, $\hat{\cM}^{data}_3(u)=port_{fix}(u,p(u))$.
Moreover, a message is sent from $v$ to $u$ containing the new
adversary port number at $v$, i.e., $port_{fix}(v,u)$. When $u$
receives this message it sets the second field of its data
sublabel to be the value $port_{fix}(v,u)$, i.e,
$\hat{\cM}^{data}_2(u)=port_{fix}(p(u),u)$.
\end{enumerate}

It follows that at all  times the index sublabel of each vertex $v$
is
$$\hat{\cM}^{index}(v)=\langle \hat{\cM}_{des}(p(v))~,~
port_{des}(p(v),v)\rangle$$ and the data sublabel of $v$ is
$$\hat{\cM}^{data}(v)=\langle \hat{\cM}_{des}(v)~,~
port_{fix}(p(v),v)~,~ port_{fix}(v,p(v)) \rangle.$$

The query and decoder algorithms of $\hat{\varphi}_{fix}$, the
resulted dynamic 1-query labeling scheme, relate to
$\hat{\pi}_{des}$ similarly to the way the query and decoder
algorithms of $\varphi_{fix}$ relate to ${\pi}_{des}$. It follows
that $\hat{\varphi}_{fix}$ is a correct dynamic 1-query routing
labeling scheme with asymptotically the same label size as
$\hat{\pi}_{des}$. Since the labels of $\pi_{des}'$ (containing
$O(\log n)$ bits) can be assigned by a distributed algorithm using
$O(n)$ messages, if $\hat{\cM}_{des}$ is given by Theorem 4.16 of
\cite{KPR04}, then the label size and amortized message complexity
of $\hat{\cM}_{des}$ are as indicated (for $\hat{\cM}_{fix}$) in
the second item of Theorem \ref{dyn}. Moreover, if
$\hat{\cM}_{des}$ is given by Theorem 1 (respectively, Theorem 2)
of \cite{K05}, then the label size and amortized message
complexity of $\hat{\cM}_{des}$ are as indicated (for
$\hat{\cM}_{fix}$) in the first (resp., third) item of Theorem
\ref{dyn}. It therefore remains to show that the number of
messages used by the marker protocol $\hat{\cM}_{fix}$ is
asymptotically the same as the number of messages used by the
marker protocol of $\hat{\cM}_{des}$. Clearly, we only need to
show that the number of messages resulted from modifying
$\hat{\cM}_{des}$ into $\hat{\cM}_{fix}$ (as described in Steps
1-5 in the description of Algorithm $\hat{\cM}_{fix}$) does not
affect the asymptotic message complexity of of $\hat{\cM}_{des}$.
Step 1 in the description of Algorithm $\hat{\cM}_{fix}$ does not
incurs any messages. Steps 3 and 5 in the description of Algorithm
$\hat{\cM}_{fix}$ incur $O(1)$ amortized message complexity per
topological change. Note that in the dynamic version of the
fixed-port model, a port is assigned a number by the adversary
only once: when a leaf $u$ joins the tree as a child of some
vertex $v$, the port number at $u$ and the port number at $v$
leading to $u$ are assigned a port number by the adversary.
Therefore, Step 4 in the description of protocol $\hat{\cM}_{fix}$
also incurs $O(1)$ amortized message complexity per topological
change. It remains to bound the number of messages incurred  by
Step 2. As mentioned before, in the translation scheme of
\cite{KPR04}, a non-leaf vertex $v$ may change its label only when
a shuffle operation is invoked on a subtree $T'$ containing
$T(v)$, the subtree containing $v$ and all of its descendants (see
Subsections 4.1.3 and 4.1.4 in \cite{KPR04}). In this shuffle
operation, a distributed algorithm assigning the labels of the
corresponding static scheme (in this case $\pi_{des}$) is invoked
on $T'$. Therefore, the number of messages incurred by this
shuffle operation is bounded from below by the number of vertices
in $T'$. It follows that if $\hat{\cM}_{des}$ is given by Theorem
4.16 of \cite{KPR04}, then the extra number of messages incurred
by Step 2 in the description of protocol $\hat{\cM}_{fix}$ is
bounded from above by the message complexity of $\hat{\pi}_{des}$.
This completes the proof of Item 2 in the theorem.

We now bound the number of messages incurred by Step 2 in the case
$\hat{\cM}_{des}$ is given by the translation method of \cite{K05}.
This is needed in order to prove Items 1 and 3 in the theorem. In
this case (similarly to the translation method of \cite{KPR04}) a
non-leaf vertex $v$ may change its label only when a reset operation
is invoked on a subtree $T'$ containing $v$. The number of messages
incurred by this reset operation is bounded from below by the number
of vertices in $T'$. Let $r'$ be the root of $T'$. In contrast to
the subtree on which the shuffle operation of \cite{KPR04} is
invoked on, the subtree $T'$ may not necessarily contain all of
$r'$'s children. In fact $T'$ is a subtree of the dynamic tree $T$
which is composed of a root vertex $r'$, a set of children
$\{u_i\}_{i\in I}$ (for some $I$ defined in \cite{KPR04}) of $r'$
and all the descendant the vertices in $\{u_i\}_{i\in I}$. We now
show that when applying the translation scheme of \cite{K05} on the
modified routing labeling scheme  $\pi'_{des}$, a non-leaf vertex
$v$ with more than one child may change its label only when a reset
operation is invoked on a subtree $T'$ containing $v$'s parent.

Note that in $\hat{\pi}_{des}$, when a vertex $v$ is added to the
tree, $v$ participates in a reset operation applied on a subtree
containing $p(v)$. Fix some vertex $v$. Let $t_o(v)$ be a time in
which $v$ participated in a reset operation applied on a subtree
containing $p(v)$ and let $L_0(v)$ be the label given to $v$ in this
reset operation. The first time after $t_0(v)$ that $v$ changed its
label, was either due to another reset operation which was invoked
on a subtree containing $p(v)$  or when a (first) child of $v$ was
added to the tree and a reset operation was applied on a subtree
rooted at $v$. In the latter case, let $L_1(v)$ be the label given
to $v$ in this reset operation, and let $t_1(v)$ be the time when
$v$ received this label. Since $\langle 1\rangle$ is the label given
by $\pi'_{des}$ to the root of any tree, it follows from the
description of Scheme $FSDL^k_p$ in \cite{K05}, that
$L_1(v)=L_0(v)\circ \langle 1\rangle$. At any time from $t_1(v)$
until the next time $v$ changed its label as a result of a reset
operation containing $p(v)$, whenever $v$ participated in a reset
operation invoked on some subtree $T'$, the subtree $T'$ must be
rooted at $v$. Moreover, the new label given to $v$ by the reset
operation on $T'$ is $L_0(v)$ concatenated with the label given to
$v$ by the static routing labeling scheme $\pi'_{des}$, which is
$\langle 1\rangle$, since $v$ is the root of $T'$. Therefore, the
label of $v$ remains $L_1(v)=L_0(v)\circ \langle 1\rangle$. It
follows that a non-leaf vertex $v$ with more than one child may
change its label only when a reset operation is invoked on a subtree
$T'$ containing $v$'s parent. Therefore, $T'$ contains all of $v$'s
children. It follows that the extra number of messages incurred by
Step 2 in the description of protocol $\hat{\cM}_{fix}$ is bounded
from above by the message complexity of $\hat{\pi}_{des}$, as
desired. Altogether, we obtain that the number of message used by
the marker protocol $\hat{\cM}_{fix}$ is asymptotically the same as
the number of message used by the marker protocol of
$\hat{\cM}_{des}$, as desired. The theorem follows. Theorems 1 and 2
of \cite{K05} imply the remaining claim of Theorem \ref{dyn}
regarding the dynamic 1-query labeling scheme supporting the routing
function.\QED

\end{document}